\documentclass[english,aps,preprint,superscriptaddress]{revtex4}
\usepackage[]{fontenc}
\usepackage[latin9]{inputenc}
\usepackage{amsmath}
\usepackage{esint}
\usepackage{epsfig}
\usepackage{graphicx}
\usepackage{bm}
\usepackage{rotating}
\usepackage{color}
\makeatletter
\@ifundefined{textcolor}{}
{%
 \definecolor{BLACK}{gray}{0}
 \definecolor{WHITE}{gray}{1}
 \definecolor{RED}{rgb}{1,0,0}
 \definecolor{GREEN}{rgb}{0,1,0}
 \definecolor{BLUE}{rgb}{0,0,1}
 \definecolor{CYAN}{cmyk}{1,0,0,0}
 \definecolor{MAGENTA}{cmyk}{0,1,0,0}
 \definecolor{YELLOW}{cmyk}{0,0,1,0}
 }

\makeatletter

\newcommand{\be}{\begin{equation}}\newcommand{\ee}{\end{equation}}\newcommand{\ba}{\begin{align}}\newcommand{\ea}{\end{align}}\def\bea{\begin{eqnarray}}\def\eea{\end{eqnarray}}

\usepackage{slashed}

\makeatother

\usepackage{babel}

\begin{document}

\title{Parallel Lepton Mass Matrices with Texture/Cofactor Zeros}

\author{Weijian Wang}

\affiliation{Department of Physics, North China Electric Power
University, Baoding 071003, P. R. China} \email{wjnwang96@gmail.com}

\begin{abstract}
In this paper we investigate the parallel texture structures
containing texture zeros in charged lepton mass matrix $M_{l}$ and
cofactor zeros in neutrino mass matrix $M_{\nu}$. These textures are
interesting since they are related to the $Z_{n}$ flavor symmetries.
Using the weak basis permutation transformation, the 15 parallel
textures are grouped as 4 classes (class I,II,III and IV) with the
matrices in each class sharing the same physical implications. Under
the current experimental data, the class I, III with inverted mass
hierarchy and class II with normal mass hierarchy are
phenomenologically acceptable. The correlations between some
important physical variables are presented, which are essential for
the model selection and can be text by future experiments. The model
realization is illustrated by means of $Z_{4}\times Z_{2}$ flavor
symmetry.

\vspace{1em}

{\textbf{PACS}}: 14.60.Pq, 12.15.Ff, 11.30.Hv
\end{abstract}

\maketitle

\section{Introduction}
The discovery of neutrino oscillations have provided us with
convincing evidences for massive neutrinos and leptonic flavor
mixing with high degree of accuracy\cite{neu1,neu2,neu3}. The
understanding of the leptonic flavor structure is one of the major
open questions in particle physics. Several attempts have been
proposed to explain the origin of neutrino mass and the observed
pattern of leptonic mixing by introducing the flavor symmetries
within the framework of seesaw models\cite{seesaw}. The flavor
symmetry often reduces the number of free parameters and leads to
the specific structures of fermion mass matrices including texture
zeros\cite{zero,zero1,GCB,z,z1}, hybrid textures\cite{hybrid,
hybrid2}, zero trace\cite{sum}, zero determinant\cite{det},
vanishing minors\cite{minor1,minor,minor2}, two traceless
submatrices\cite{tra}, equal elements or cofactors\cite{co}, inverse
hybrid textures\cite{hyco}. Among these models, the matrices with
texture or cofactor zeros are particularly interesting due to their
connections to the flavor symmetries. The phenomenological
examination of texture zeros or cofactor zeros in flavor basis have
been widely studies in Ref.\cite{zero,zero1,minor1,minor,minor2}
where the charged lepton mass matrices $M_{l}$ are diagonal.
However, no universal principle is required that the flavor basis is
necessary and the more general cases should be considered in no
diagonal $M_{l}$ basis. In this scenario, the lepton mass matrices
with texture zeros in both lepton mass $M_{l}$ and neutrino mass
matrix $M_{\nu}$ have been systematically investigated by many
authors\cite{GCB,z}(for a review, see \cite{z1}).

In this paper, we propose the new possible texture structures where
there are two texture zeros in $M_{l}$ and two cofactor zeros in
$M_{\nu}$ (We denote them the matrices with texture/cofactor zeros).
It seems that such mass matrices are rather unusual because one
instinctively expects the type of texture structures to be the same
for both $M_{l}$ and $M_{\nu}$. However, one reminds the type-I
seesaw model as $ M_{\nu}=-M_{D} M_{R}^{-1}M_{D}^{T}$. Then the
texture or cofactor zeros of $M_{\nu}$ can be attributed to the
texture zeros in $M_{D}$ and $M_{R}$. Generally, this can be
realized by $Z_{n}$ flavor symmetry\cite{zn,minor1}. Therefore from
the point of flavor symmetry, both texture zeros and cofactor zeros
structures manifest the same flavor symmetry in different ways. It
is our main motivation to carry out this work and a concrete model
will be constructed in the following section. Furthermore, we take
the so-called the parallel $Ans\ddot{a}tze$ that the positions of
texture zeros in $M_{l}$ are chosen to be the same as the cofactor
zeros in $M_{\nu}$. Although there is no priori reason requiring the
parallel structures, they are usually regarded in many literatures
as an esthetical appeal and the precursor of the more general cases.
The lepton mass matrices with parallel texture zero structures have
been systematically investigated in Ref.\cite{GCB}. Subsequently,
the idea is generalized to more complicated situations such as
parallel hybrid textures\cite{hyp}, parallel cofactor zero
textures\cite{myco}. In our case, there exists $C^{2}_{6}=15$
logically possible patterns for two texture/cofactor zeros in mass
matrices. It is indicated that the 15 textures can be grouped into 4
classes with the matrices in each class connected by $S_{3}$
permutation transformation and sharing the same physical
implications. Among the 4 classes, one of them is not viable
phenomenologically. Therefore we focus on the other three nontrivial
classes.

The paper is organized as follow. In Sec. II, we present the
classification of mass matrices and relate them to the current
experimental results. In Sec. III, we diagonalize the mass matrices,
confront the numerical results with the experimental data and
discuss their predictions. In Sec. IV, the model realization is
given under the $Z_{4}\times Z_{2}$ flavor symmetry. We summarize
the results in Sec. V.

\section{Formalism}
\subsection{Weak basis equivalent classes}
As shown in Ref.\cite{GCB}, there exists the general weak basis (WB)
transformations leaving gauge currents invariant i.e
\begin{equation}
M_{l}\rightarrow M_{l}^{\prime}=W^{\dagger}M_{l}W_{R}\quad\quad\quad
M_{\nu}\rightarrow M_{\nu}^{\prime}=W^{T}M_{\nu}W
\end{equation}
where the neutrinos are assumed to be Majorana fermions and $W$,
$W_{R}$ are $3\times 3$ unitary matrices. Two matrices related by WB
transformations have the same physical implications. Therefore the
parallel matrices with texture/cofactor zeros located at different
positions can be connected by $S_{3}$ permutation matrix $P$ as a
specific WB transformation
\begin{equation}
M_{l}^{\prime}=P^{T}M_{l}P\quad\quad\quad
M_{\nu}^{\prime}=P^{T}M_{\nu}P
\end{equation}
It is noted that $P$ changes the positions of cofactor zero elements
but still preserves the parallel structures for both charged lepton
and neutrino mass textures. Then the texture/cofactor zeros matrices
are classified into 4 classes:

Class I:
\begin{equation}\begin{split}
\left(\begin{array}{ccc}
0/\bigtriangleup&\times&0/\bigtriangleup\\
\times&\times&\times\\
0/\bigtriangleup&\times&\times
\end{array}\right)\quad\quad
\left(\begin{array}{ccc}
0/\bigtriangleup&0/\bigtriangleup&\times\\
0/\bigtriangleup&\times&\times\\
\times&\times&\times
\end{array}\right)\quad\quad
\left(\begin{array}{ccc}
\times&0/\bigtriangleup&\times\\
0/\bigtriangleup&0/\bigtriangleup&\times\\
\times&\times&\times
\end{array}\right)\\
\left(\begin{array}{ccc}
\times&\times&\times\\
\times&0/\bigtriangleup&0/\bigtriangleup\\
\times&0/\bigtriangleup&\times
\end{array}\right)\quad\quad
\left(\begin{array}{ccc}
\times&\times&0/\bigtriangleup\\
\times&\times&\times\\
0/\bigtriangleup&\times&0/\bigtriangleup
\end{array}\right)\quad\quad
\left(\begin{array}{ccc}
\times&\times&\times\\
\times&\times&0/\bigtriangleup\\
\times&0/\bigtriangleup&0/\bigtriangleup
\end{array}\right)
\end{split}\label{matrx}\end{equation}

Class II:
\begin{equation}\begin{split}
\left(\begin{array}{ccc}
0/\bigtriangleup&\times&\times\\
\times&\times&0/\bigtriangleup\\
\times&0/\bigtriangleup&\times
\end{array}\right)\quad\quad
\left(\begin{array}{ccc}
\times&\times&0/\bigtriangleup\\
\times&0/\bigtriangleup&\times\\
0/\bigtriangleup&\times&\times
\end{array}\right)\quad\quad
\left(\begin{array}{ccc}
\times&0/\bigtriangleup&\times\\
0/\bigtriangleup&\times&\times\\
\times&\times&0/\bigtriangleup
\end{array}\right)
\end{split}\label{matrx}\end{equation}

Class III:
\begin{equation}\begin{split}
\left(\begin{array}{ccc}
0/\bigtriangleup&\times&\times\\
\times&0/\bigtriangleup&\times\\
\times&\times&\times
\end{array}\right)\quad\quad
\left(\begin{array}{ccc}
0/\bigtriangleup&\times&\times\\
\times&\times&\times\\
\times&\times&0/\bigtriangleup
\end{array}\right)\quad\quad
\left(\begin{array}{ccc}
\times&\times&\times\\
\times&0/\bigtriangleup&\times\\
\times&\times&0/\bigtriangleup
\end{array}\right)
\end{split}\label{matrx}\end{equation}

Class IV:
\begin{equation}\begin{split}
\left(\begin{array}{ccc}
\times&0/\bigtriangleup&0/\bigtriangleup\\
0/\bigtriangleup&\times&\times\\
0/\bigtriangleup&\times&\times
\end{array}\right)\quad\quad
\left(\begin{array}{ccc}
\times&0/\bigtriangleup&\times\\
0/\bigtriangleup&\times&0/\bigtriangleup\\
\times&0/\bigtriangleup&\times
\end{array}\right)\quad\quad
\left(\begin{array}{ccc}
\times&\times&0/\bigtriangleup\\
\times&\times&0/\bigtriangleup\\
0/\bigtriangleup&0/\bigtriangleup&\times
\end{array}\right)
\end{split}\label{matrx}\end{equation}
where "$0/\bigtriangleup$" at $(i,j)$ position represents the
texture zero condition $M_{ij}=0$ and the cofactor zero condition
$C_{ij}=0$; The "$\times$" denotes arbitrary element. One can check
that the matrices with cofactor zeros in class I are equivalent to
the texture zero ones. Choosing the first matrix of class I as an
example, we have
\begin{equation}\begin{split}
M_{\nu}=\left(\begin{array}{ccc}
\bigtriangleup&\times&\bigtriangleup\\
\times&\times&\times\\
\bigtriangleup&\times&\times
\end{array}\right)\Rightarrow
M_{\nu}^{-1}=\left(\begin{array}{ccc}
0&\times&0\\
\times&\times&\times\\
0&\times&\times
\end{array}\right)
\Rightarrow M_{\nu}=\left(\begin{array}{ccc}
\times&\times&\times\\
\times&0&0\\
\times&0&\times
\end{array}\right)
\end{split}\label{matrx1}\end{equation}
Thus the parallel texture structures of class I are equivalent to
the no-parallel structures with two texture zeros. Although the
parallel texture zero structures has been explored
extensively\cite{GCB,z,z1}, the analysis of the no-parallel two
texture zero structure has not yet been reported. On the other hand,
as having been pointed out in Ref.\cite{GCB,myco}, the class IV
leads to the decoupling of a generation of lepton from mixing and
thus not experimentally viable.

\subsection{Useful notations}
As we have mentioned, among the 4 classes only class I, II and III
are nontrivial. We represent them as
\begin{equation}\begin{split}
M_{l/\nu}^{I}=\left(\begin{array}{ccc}
0/\bigtriangleup&\times&0\bigtriangleup\\
\times&\times&\times\\
0\bigtriangleup&\times&\times
\end{array}\right)\quad\quad
M_{l/\nu}^{II}=\left(\begin{array}{ccc}
0/\bigtriangleup&\times&\times\\
\times&\times&0/\bigtriangleup\\
\times&0/\bigtriangleup&\times
\end{array}\right)\quad\quad
M_{l/\nu}^{III}=\left(\begin{array}{ccc}
0/\bigtriangleup&\times&\times\\
\times&0/\bigtriangleup&\times\\
\times&\times&\times
\end{array}\right)
\end{split}\label{matrx}\end{equation}
In the analysis, we consider $M_{l}$ is to be Hermitian and the
Majorana neutrino mass texture $M_{\nu}$ is complex and symmetric.
The $M_{l}$ and $M_{\nu}$ are diagonalized by unitary matrix $V_{l}$
and $V_{\nu}$
\begin{equation}
M_{l}=V_{l}M_{l}^{D}V_{l}^{\dagger}\quad\quad
M_{\nu}=V_{\nu}M_{\nu}^{D}V_{\nu}^{T}
\label{dia}\end{equation}
where $M_{l}^{D}=Diag(m_{e},m_{\mu}, m_{\tau})$,
$M_{\nu}^{D}=Diag(m_{1},m_{2},m_{3})$. The
Pontecorvo-Maki-Nakagawa-Sakata matrix\cite{PMNS} $U_{PMNS}$ is
given by
\begin{equation}
U_{PMNS}=V_{l}^{\dagger}V_{\nu} \label{upmns}\end{equation} and
parameterized as
\begin{equation}
U_{PMNS}=UP_{\nu}=\left(\begin{array}{ccc}
c_{12}c_{13}&c_{13}s_{12}&s_{13}e^{-i\delta}\\
-s_{12}c_{23}-c_{12}s_{13}s_{23}e^{i\delta}&c_{12}c_{23}-s_{12}s_{13}s_{23}e^{i\delta}&c_{13}s_{23}\\
s_{23}s_{12}-c_{12}c_{23}s_{13}e^{i\delta}&-c_{12}s_{23}-c_{23}s_{12}s_{13}e^{i\delta}&c_{13}c_{23}
\end{array}\right)\left(\begin{array}{ccc}
1&0&0\\
0&e^{i\alpha}&0\\
0&0&e^{i(\beta+\delta)}
\end{array}\right)
\label{3}\end{equation} where we use the abbreviation
$s_{ij}=\sin\theta_{ij}$ and $c_{ij}=\cos\theta_{ij}$. The
($\alpha$,$\beta$) in $P_{\nu}$ represents the two Majorana
CP-violating phases and $\delta$ denotes the Dirac CP-violating
phase. In order to facilitate our calculation, we treat the
Hermitian matrix $M_{l}$ factorisable. i.e
\begin{equation}
M_{l}=K_{l}M_{l}^{r}K_{l}^{\dagger}
\end{equation}
where $K_{l}$ is the unitary phase matrix parameterized as
$K_{l}=diag(1,e^{i\phi_{1}},e^{i\phi_{2}})$. The $M_{l}^{r}$ becomes
a real symmetric matrix which can be diagonalized by real orthogonal
matrix $O_{l}$. Then we have
\begin{equation}
V_{l}=K_{l}O_{l}
\end{equation}
and
\begin{equation}
U_{PMNS}=O_{l}^{T}K_{l}^{\dagger}V_{\nu} \label{okv}\end{equation}
From \eqref{dia}, \eqref{upmns} and \eqref{okv}, the neutrino mass
matrix $M_{\nu}$ is given by
\begin{equation}
M_{\nu}=K_{l}VP_{\nu}M_{\nu}^{D}P_{\nu}V^{T}K_{l}^{\dagger}
\label{mvd}\end{equation} where $V\equiv O_{l}U$. From \eqref{mvd}
and solving the cofactor zero conditions of $M_{\nu}$
\begin{equation}
M_{\nu(pq)}M_{\nu(rs)}-M_{\nu(tu)}M_{\nu(vw)}=0\quad\quad
M_{\nu(p^{\prime}q^{\prime})}M_{\nu(r^{\prime}s^{\prime})}-M_{\nu(t^{\prime}u^{\prime})}M_{\nu(v^{\prime}w^{\prime})}=0
\label{czc}\end{equation} we get
\begin{equation}
\frac{m_{1}}{m_{2}}e^{-2i\alpha}=\frac{K_{3}L_{1}-K_{1}L_{3}}{K_{2}L_{3}-K_{3}L_{2}}
\label{r1}\end{equation}
\begin{equation}
\frac{m_{1}}{m_{3}}e^{-2i\beta}=\frac{K_{2}L_{1}-K_{1}L_{2}}{K_{3}L_{2}-K_{2}L_{3}}e^{2i\delta}
\label{r2}\end{equation} where
\begin{equation}
K_{i}=(V_{pj}V_{qj}V_{rk}V_{sk}-V_{tj}V_{uj}V_{vk}V_{wk})+(j\leftrightarrow
k)\end{equation}
\begin{equation}
L_{i}=(V_{p^{\prime}j}V_{q^{\prime}j}V_{r^{\prime}k}V_{s^{\prime}k}-V_{t^{\prime}j}V_{u^{\prime}j}V_{v^{\prime}k}V_{w^{\prime}k})+(j\leftrightarrow
k)\end{equation} with $(i,j,k)$ a cyclic permutation of (1,2,3).
With the help of Eq.\eqref{r1} and \eqref{r2}, the magnitudes of
neutrino mass radios are given by
\begin{equation}
\rho=\Big|\frac{m_{1}}{m_{3}}e^{-2i\beta}\Big|
\label{t1}\end{equation}
\begin{equation}
\sigma=\Big|\frac{m_{1}}{m_{2}}e^{-2i\alpha}\Big|
\label{t2}\end{equation} with the two Majorana CP-violating phases
\begin{equation}
\alpha=-\frac{1}{2}arg\Big(\frac{K_{3}L_{1}-K_{1}L_{3}}{K_{2}L_{3}-K_{3}L_{2}}\Big)
\label{21}\end{equation}
\begin{equation}
\beta=-\frac{1}{2}arg\Big(\frac{K_{2}L_{1}-K_{1}L_{2}}{K_{3}L_{3}-K_{2}L_{3}}e^{2i\delta}\Big)
\label{22}\end{equation} The results of Eq. \eqref{t1},\eqref{t2},
\eqref{21} and \eqref{22} imply that the two mass ratio ($\rho$ and
$\sigma$) and two Majorana CP-violating phases ($\alpha$ and
$\beta$) are fully determined in terms of the real orthogonal matrix
$O_{l}$, $U$($\theta_{12}, \theta_{23}, \theta_{13}$ and $\delta$).
The neutrino mass ratios $\rho$ and $\sigma$ are related to the
ratio of two neutrino mass-squared differences defined as
\begin{equation}
R_{\nu}\equiv\frac{\delta m^{2}}{\Delta
m^{2}}=\frac{2\rho^{2}(1-\sigma^{2})}{|2\sigma^{2}-\rho^{2}-\rho^{2}\sigma^{2}|}
\label{rv}\end{equation} where $\delta m^{2}\equiv
m_{2}^{2}-m_{1}^{2}$ and $\Delta m^{2}\equiv \mid
m_{3}^{2}-\frac{1}{2}(m_{1}^{2}+m_{2}^{2})\mid$. The three neutrino
mass eigenvalues $m_{1},m_{2}$ and $m_{3}$ are given by
\begin{equation}
 m_{2}=\sqrt{\frac{\delta m^{2}}{1-\sigma^{2}}}\quad\quad
 m_{1}=\sigma m_{2}\quad\quad m_{3}=\frac{m_{1}}{\rho}
\label{abm}\end{equation}  In the following numerical analysis, we
utilize the recent 3$\sigma$ confidential level global-fit data from
the neutrino oscillation experiments\cite{data}.i.e
\begin{equation}\begin{split}
\sin^{2}\theta_{12}/10^{-1}=3.08^{+0.51}_{-0.49}\quad
\sin^{2}\theta_{23}/10^{-1}=4.25^{+2.16}_{-0.68}\quad
\sin^{2}\theta_{13}/10^{-2}=2.34^{+0.63}_{-0.57}\\
\delta m^{2}/10^{-5}=7.54^{+0.64}_{-0.55} eV^{2}\quad\quad
\bigtriangleup m^{2}/10^{-3}=2.44^{+0.22}_{-0.22} eV^{2}
\end{split}\end{equation}
for normal hierarchy (NH) and
\begin{equation}\begin{split}
\sin^{2}\theta_{12}/10^{-1}=3.08^{+0.51}_{-0.49}\quad
\sin^{2}\theta_{23}/10^{-1}=4.25^{+2.22}_{-0.74}\quad
\sin^{2}\theta_{13}/10^{-2}=2.34^{+0.61}_{-0.61}\\
\delta m^{2}/10^{-5}=7.54^{+0.64}_{-0.55} eV^{2}\quad\quad
\bigtriangleup m^{2}/10^{-3}=2.40^{+0.21}_{-0.23} eV^{2}
\end{split}\end{equation}
for inverted hierarchy(IH). By this time, no constraint is added on
the Dirac CP-violating phase $\delta$ at $3\sigma$ level, however
the recent numerical analysis\cite{data} tends to give the best-fit
value $\delta\approx 1.40\pi$. In neutrino oscillation experiments,
the CP violation effect is usually reflected by the Jarlskog
rephasing invariant quantity\cite{Jas} defined as
\begin{equation}
J_{CP}=s_{12}s_{23}s_{13}c_{12}c_{23}c_{13}^{2}\sin\delta
\end{equation}
The Majorana nature of neutrino can be determined if any signal of
neutrinoless double decay($0\nu\beta\beta$) is observed, implying
the violation of leptonic number violation. The decay ratio is
related to the effective Majorana neutrino mass $m_{ee}$, which is
written as
\begin{equation}
m_{ee}=|m_{1}c_{12}^{2}c_{13}^{2}+m_{2}s_{12}^{2}c_{13}^{2}e^{2i\alpha}+m_{3}s_{13}^{2}e^{2i\beta}|
\end{equation}
 Although a $3\sigma$ result of $m_{ee}=(0.11-0.56)$ eV is
reported by the Heidelberg-Moscow Collaboration\cite{HM}, this
result is criticized\cite{NND2} and shall be checked by the
forthcoming experiment. It is believed that that the next generation
$0\nu\beta\beta$ experiments, with the sensitivity of $ m_{ee}$
being up to 0.01 eV\cite{NDD}, will open the window to not only the
absolute neutrino mass scale but also the Majorana-type CP
violation. Besides the $0\nu\beta\beta$ experiments, a more severe
constraint was set from the recent cosmology observation. Recently,
an upper bound on the sum of neutrino mass $\sum m_{i}<0.23$ eV is
reported by Plank Collaboration\cite{Planck} combined with the WMAP,
high-resolution CMB and BAO experiments.

\section{Numerical analysis}

We have proposed a detailed numerical analysis for class I, II and
III. In this section we presented the main predictions of all the
classes.

\subsection{Class I}
Let's start from the factorisable formation of charged lepton matrix
$M_{l}^{r}$
\begin{equation}\begin{split}
(M_{l}^{r})^{I}=\left(\begin{array}{ccc}
0&a&0\\
a&b&c\\
0&c&d
\end{array}\right)
\end{split}\label{I}\end{equation}
As proposed in Ref.\cite{GCB, myco}, the coefficients $a, b$ and $c$
are assumed to be real and positive without losing generality. The
real coefficient $d$ is treated as a free parameter. Then the matrix
\eqref{I} can be diagonalized by an orthogonal matrix $O_{l}$
\begin{equation}
O_{l}^{T}(M_{l}^{r})^{I}O_{l}=diag(m_{e},- m_{\mu},m_{\tau})
\label{ot2}\end{equation} where the minus sign in \eqref{ot2} is
introduced to facilitate the analytical calculation and has no
physical meaning since it originates from the phase transformation
of Dirac fermions. Following the same strategy of Ref.\cite{GCB} and
using the invariant Tr$(M_{l}^{r})$, Det$ (M_{l}^{r})$ and
Tr$(M_{l}^{r})^{2}$, the nozero elements of $M_{l}^{r}$ can be
expressed in terms of three mass eigenvalues $m_{e}, m_{\mu}$,
$m_{\tau}$ and $d$
\begin{equation}
a=\sqrt{\frac{m_{e}m_{\mu}m_{\tau}}{d}} \label{IA}\end{equation}
\begin{equation}
b=m_{e}-m_{\mu}+m_{\tau}-d \label{IB}\end{equation}
\begin{equation}
c=\sqrt{-\frac{(d-m_{e})(d+m_{\mu})(d-m_{\tau})}{d}}
\label{IC}\end{equation} Using the expression \eqref{IA}, \eqref{IB}
and \eqref{IC}, $O_{l}$ can be constructed. Here we adopt the result
of \cite{GCB} i.e
\begin{equation}\begin{split}
O_{l}=\left(\begin{array}{ccc}
\sqrt{\frac{m_{\mu}m_{\tau}(d-m_{e})}{d(m_{\mu}+m_{e})(m_{\tau}-m_{e})}}&\sqrt{\frac{m_{e}m_{\tau}(m_{\mu}+d)}{d(m_{\mu}+m_{e})(m_{\tau}+m_{\mu})}}&\sqrt{-\frac{m_{e}m_{\mu}(d-m_{\tau})}{d(m_{\tau}-m_{e})(m_{\tau}+m_{\mu})}}\\
\sqrt{-\frac{m_{e}(m_{e}-d)}{(m_{\mu}+m_{e})(m_{\tau}-m_{e})}}&-\sqrt{-\frac{m_{\mu}(d+m_{\mu})}{(m_{\mu}+m_{e})(m_{\tau}+m_{\mu})}}&\sqrt{\frac{m_{\tau}(m_{\tau}-d)}{(m_{\tau}-m_{e})(m_{\tau}+m_{\mu})}}\\
-\sqrt{-\frac{m_{e}(d+m_{\mu})(d-m_{\tau})}{d(m_{\mu}+m_{e})(m_{\tau}-m_{e})}}&\sqrt{\frac{m_{\mu}(d-m_{e})(m_{\tau}-d)}{d(m_{\mu}+m_{e})(m_{\tau}-m_{e})}}&\sqrt{\frac{m_{\tau}(d-m_{e})(d+m_{\mu})}{d(m_{\tau}-m_{e})(m_{\tau}+m_{\mu})}}
\end{array}\right)
\end{split}\label{II1}\end{equation}
Replacing the \eqref{t1}, \eqref{t2}, \eqref{21}, \eqref{22} and
\eqref{rv} with the $O_{l}$ obtained in \eqref{II1}, we can see that
the ratios of mass ($\rho, \sigma$), two Majorana CP-violating
phases $(\alpha, \beta)$ and the ratio of mass squared difference
$R_{\nu}$ can be expressed via eight parameters: three mixing angle
$\theta_{12}, \theta_{23}, \theta_{13}$, one Dirac CP violating
phase $\delta$, three charged lepton mass $(m_{e}, m_{\mu},
m_{\tau})$ and the parameter $d$. Here we choose the three charged
lepton mass at the electroweak scale($\mu\simeq M_{Z}$)
i.e\cite{zzx2}
\begin{equation}
m_{e}=0.486570154 MeV\quad\quad m_{\mu}=102.7181377MeV \quad\quad
m_{\tau}=1746.17MeV
\end{equation}

\begin{figure}
 \includegraphics[scale = 0.7]{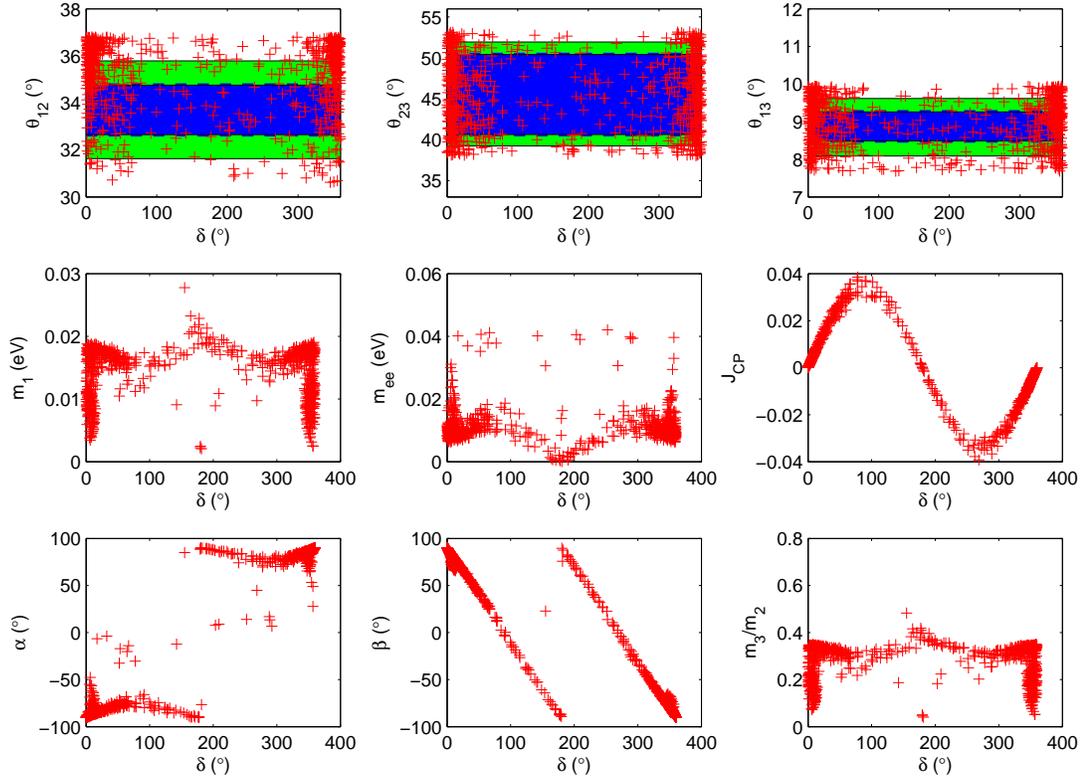}%
\caption {The correlation plots for class I(IH). The blue horizontal
bands represent the 1$\sigma$ uncertainty in determination of
$\theta_{12}, \theta_{23}$ and $\theta_{13}$ while they plus the
green horizontal bands correspond to the 2$\sigma$ uncertainty.}
\label{IIH}
\end{figure}
In the numerical analysis, a set of random numbers are generated for
the three mixing angles $(\theta_{12}, \theta_{23}, \theta_{13})$
and mass square differences ($\delta m^{2}, \Delta m^{2}$) in their
$3\sigma$ range. We also randomly vary the parameter $d$ in its
appropriate range. Since at 3 $\sigma$ level the Dirac CP-violating
phase $\delta$ is unconstrained in neutrino oscillation experiments,
we vary it randomly in the range of $[0,2\pi)$. With the random
number and using Eq. \eqref{t1}, \eqref{t2} and \eqref{rv}, neutrino
mass ratios $(\rho,\sigma)$ and the mass-squared difference ratio
$R_{\nu}$ are determined. Then the input parameters is empirically
acceptable when the $R_{\nu}$ falls inside the the $3\sigma$ range
of experimental data, otherwise they are ruled out. Finally, we get
the value of neutrino mass and Majorana CP-violating $\alpha$ and
$\beta$ though Eq.\eqref{21}, \eqref{22} and \eqref{abm}. Once the
the absolute neutrino mass $m_{1,2,3}$ are obtained , the further
constraint from cosmology should be considered. In this paper, the
upper bound on the sum of neutrino mass $\Sigma m_{i}$ is set to be
less than 0.23 eV. It turns out that class I are phenomenologically
acceptable only for inverted mass hierarchy.
\begin{figure}
 \includegraphics[scale = 0.6]{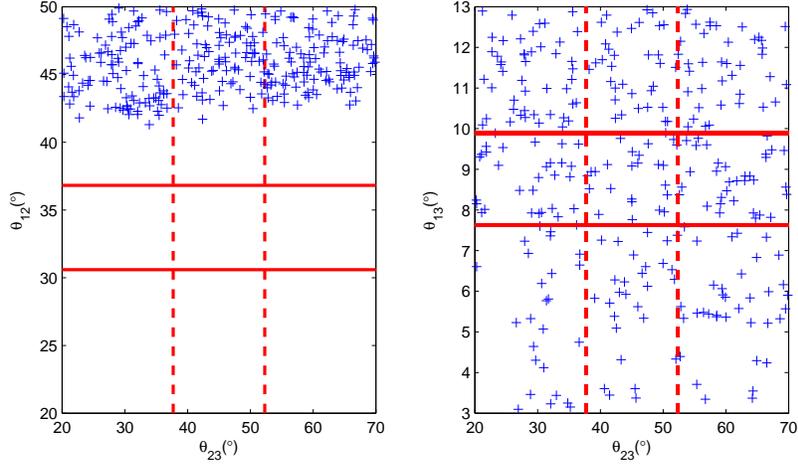}%
\caption {The correlation plots $(\theta_{23}, \theta_{12})$ and
$(\theta_{23}, \theta_{13})$ for class I(NH). The horizontal and
vertical lines respectively denote the 3$\sigma$ upper and lower
bound of $\theta_{12}$ and $\theta_{23}$} \label{INH}
\end{figure}
The predictions of class I with inverted mass hierarchy are
presented in Fig.\ref{IIH}. From the diagrams, one can see that the
three neutrino mixing angles $\theta_{12}$, $\theta_{23}$ and
$\theta_{13}$ fully cover their $3\sigma$ experimental data.
Although there is no bound on the Dirac CP-violating phase $\delta$,
a numerical preference appears at around $0^{\circ}\sim
50^{\circ}(360^{\circ}\sim 310^{\circ})$. The unrestricted $\delta$
leads to the $J_{CP}$ varying in the range of $0\sim 0.04$. There
also exists a strong correlation between $\delta$ and the lightest
neutrino mass $m_{3}$. Especially, the range
$0.002$eV$<m_{3}<0.02$eV is derived for $\delta$ lying around
$0^{\circ}(360^{\circ})$, indicating that both strong and mild mass
hierarchy are allowed. On the other hand, the mild mass hierarchy is
much more appealing for $100^{\circ}<\delta<260^{\circ}$. Although
both Majorana CP-violating phase $\alpha$ and $\beta$ is allowed in
the range of $-90^{\circ}\sim 90^{\circ}$, there shows a preferable
distribution for $\alpha$ in $\pm90^{\circ}\sim\pm50^{\circ}$ and a
strong correlation between $\delta$ and $\beta$. There exists an
upper bound of $0.05$eV on the effective Majorana neutrino mass
$m_{ee}$, leaving the possible space for detecting in future
neutrinoless double beta decay $(0\nu\beta\beta)$ experiments.

The class I with inverted hierarchy is ruled out by $3\sigma$ data.
To see this, we show the correlated plots $(\theta_{23},
\theta_{12})$ and $(\theta_{23}, \theta_{13})$ in Fig.\ref{INH}.
From the diagrams, one can see that even though $\theta_{13}$ fully
covers its $3\sigma$ range, the common parameter spaces
$(\theta_{23},\theta_{12})$ fails to provide a allowed region to
saturate the experimental constraint. Moveover, one always obtains
$\theta_{23}>40^{\circ}$, which means a rather large correction of
$\theta_{12}$ is needed to reconcile the observed PMNS matrix.

\subsection{Class II}
The factorisable formation of charged lepton matrix of class I is
given by expression:
\begin{equation}\begin{split}
(M_{l}^{r})^{II}=\left(\begin{array}{ccc}
0&a&c\\
a&b&0\\
c&0&d
\end{array}\right)
\end{split}\label{II}\end{equation}
It can be diagonalized by an orthogonal matrix $O_{l}$
\begin{equation}
O_{l}^{T}(M_{l}^{r})^{II}O_{l}=diag(m_{e},- m_{\mu},m_{\tau})
\label{ot1}\end{equation} Without losing generality, the
coefficients $a, c, d$ are set to be real and positive. Using the
invariant Tr$(M_{l}^{r})$, Det$ (M_{l}^{r})$ and
Tr$(M_{l}^{r})^{2}$, the nozero elements of $M_{l}^{r}$ are
expressed as
\begin{equation}
a=\sqrt{-\frac{(m_{e}-m_{\mu}-d)(m_{e}+m_{\tau}-d)(-m_{\mu}+m_{\tau}-d)}{m_{e}-m_{\mu}+m_{\tau}-2d}}
\label{IIA}\end{equation}
\begin{equation}
b=m_{e}-m_{\mu}+m_{\tau}-d \label{IIB}\end{equation}
\begin{equation}
c=\sqrt{\frac{(d-m_{e})(d+m_{\mu})(d-m_{\tau})}{m_{e}-m_{\mu}+m_{\tau}-2d}}
\label{IIC}\end{equation} where the parameter $d$ is allowed in the
range of $m_{e}-m_{\mu}<d<m_{e}$ and $m_{\tau}-m_{\mu}<d<m_{\tau}$.
Then the $O_{l}$ can be easily constructed as
\begin{equation}\begin{split}
O_{l}=\left(\begin{array}{ccc}
\frac{(b-m_{e})(d-m_{e})}{N_{1}}&\frac{(b+m_{\mu})(d+m_{\mu})}{N_{2}}&\frac{(b-m_{\tau})(d-m_{\tau})}{N_{3}}\\
-\frac{a(d-m_{e})}{N_{1}}&-\frac{a(d+m_{\mu})}{N_{2}}&-\frac{a(d-m_{\tau})}{N_{3}}\\
-\frac{c(b-m_{e})}{N_{3}}&-\frac{c(b+m_{\mu})}{N_{3}}&-\frac{c(b-m_{\tau})}{N_{3}}
\end{array}\right)
\end{split}\label{II1}\end{equation}
where $N_{1}$, $N_{2}$ and $N_{3}$ are the normalized coefficients
given by
\begin{equation}
N_{1}^{2}=(b-m_{e})^{2}(d-m_{e})^{2}+a^{2}(d-m_{e})^{2}+c^{2}(b-m_{\tau})^{2}
\end{equation}
\begin{equation}
N_{2}^{2}=(b+m_{\mu})^{2}(d+m_{\mu})^{2}+a^{2}(d+m_{\mu})^{2}+c^{2}(b+m_{\mu})^{2}
\end{equation}
\begin{equation}
N_{3}^{2}=(b-m_{\tau})^{2}(d-m_{\tau})^{2}+a^{2}(d-m_{\tau})^{2}+c^{2}(b-m_{\tau})^{2}
\end{equation}

\begin{figure}
 \includegraphics[scale = 0.7]{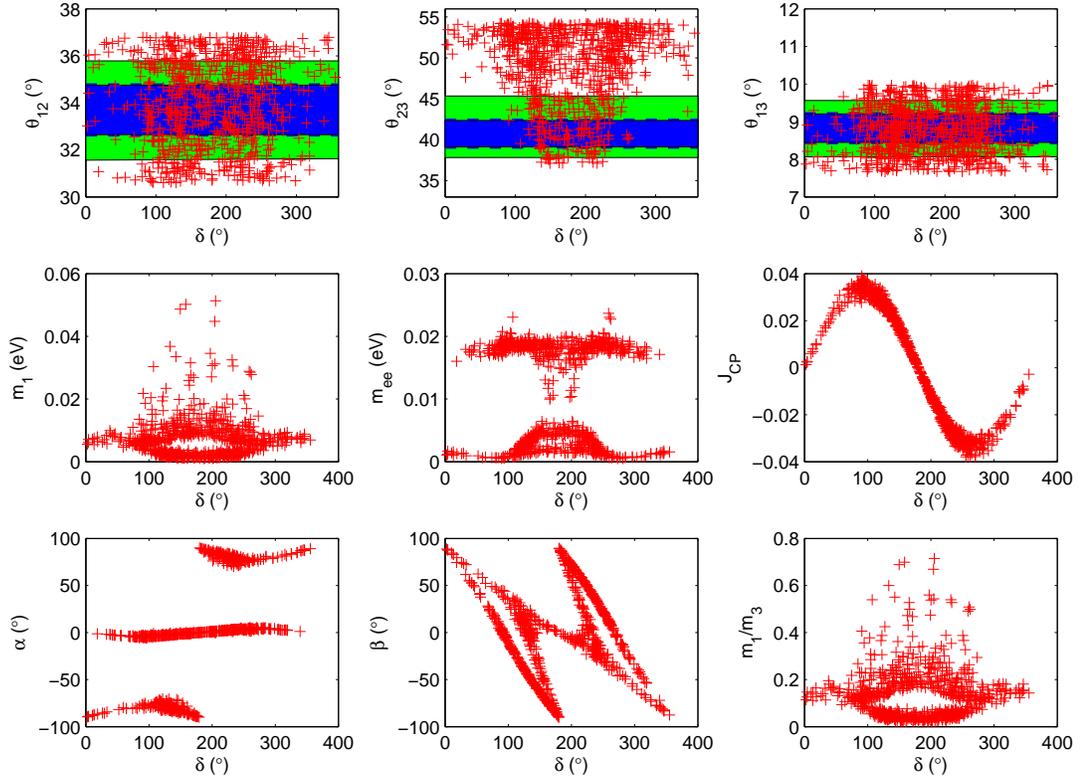}%
\caption {The correlation plots for class II(NH). The blue
horizontal bands represent the 1$\sigma$ uncertainty in
determination of $\theta_{12}, \theta_{23}$ and $\theta_{13}$ while
they plus the green horizontal bands correspond to the 2$\sigma$
uncertainty.} \label{IINH}
 \end{figure}
The numerical results of class II for normal hierarchy are presented
in Fig.\ref{IINH}. We can see from the figures that the three
neutrino mixing angle $\theta_{12}$, $\theta_{23}$ $\theta_{13}$ and
Dirac CP-violating phase $\delta$ vary arbitrarily in its $3\sigma$
range. There exhibits a strong correlation between $\delta$ and
$\theta_{23}$. Only when $\delta$ is located in the range of
$100^{\circ}\sim 260^{\circ}$, the $\theta_{23}$ has the possibility
to be less then $45^{\circ}$. This is particularly interesting since
the recent global fit trends to give the $\theta_{23}<45^{\circ}$ at
2$\sigma$ level. The strong $\delta-\theta_{23} $ correlation is
essential for the model selection and will be confirmed or ruled out
by future long-baseline neutrino oscillation experiments. The
similar correlations also holds for $\delta$, $m_{ee}$ and the
lightest neutrino mass $m_{1}$. Moveover, there exists a constrained
range of $0$eV$<m_{1}<$$0.06$eV, indicating that both strong and
mild neutrino mass hierarchy are possible. There are strong
correlations between $\alpha$, $\beta$ and $\delta$. Especially, the
Majorana CP-violating phase $\alpha$ is restricted in the range of
$-5^{\circ}\sim+5^{\circ}$ and $\pm90^{\circ}\sim\pm50^{\circ}$. The
effective Majorana neutrino mass $m_{ee}$ is highly constrained in
the two ranges of $0$eV$\sim0.008$eV and $0.01$eV$\sim0.025$eV. The
later reaches the accuracy of the future neutrinoless double beta
decay $(0\nu\beta\beta)$ experiments. We also observed that the
allowed range of Jarlskog rephasing invariant $|J_{CP}|$ is $0\sim
0.04$, which is potentially detected by future long-baseline
neutrino oscillation experiments.

\begin{figure}
 \includegraphics[scale = 0.6]{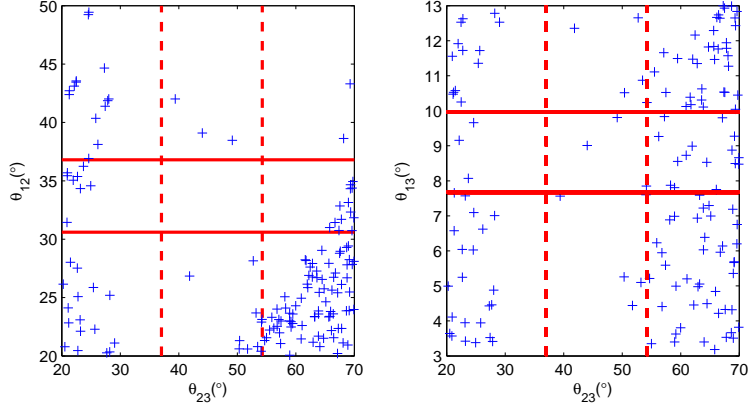}%
\caption {The correlation plots $(\theta_{23}, \theta_{12})$ and
$(\theta_{23}, \theta_{13})$ for class II(IH). The horizontal and
vertical lines respectively denote the 3$\sigma$ upper and lower
bound of $\theta_{12}$ and $\theta_{23}$} \label{IIIH}
\end{figure}

The IH case, as we can see from Fig.\ref{IIIH}, is
phenomenologically ruled by $3\sigma$ experimental data. As class I,
the theoretical prediction of $(\theta_{23}, \theta_{12})$ common
space fails to be located in its experimental region. Moreover, the
possibility distribution of $\theta_{23}$ shows a strong preference
of $\theta_{23}<33^{\circ}$ or $\theta_{23}>50^{\circ}$, which means
a large correction of $\theta_{23}$ angle is needed to produce the
$2\sigma$ global-fit value.

\begin{figure}
 \includegraphics[scale = 0.7]{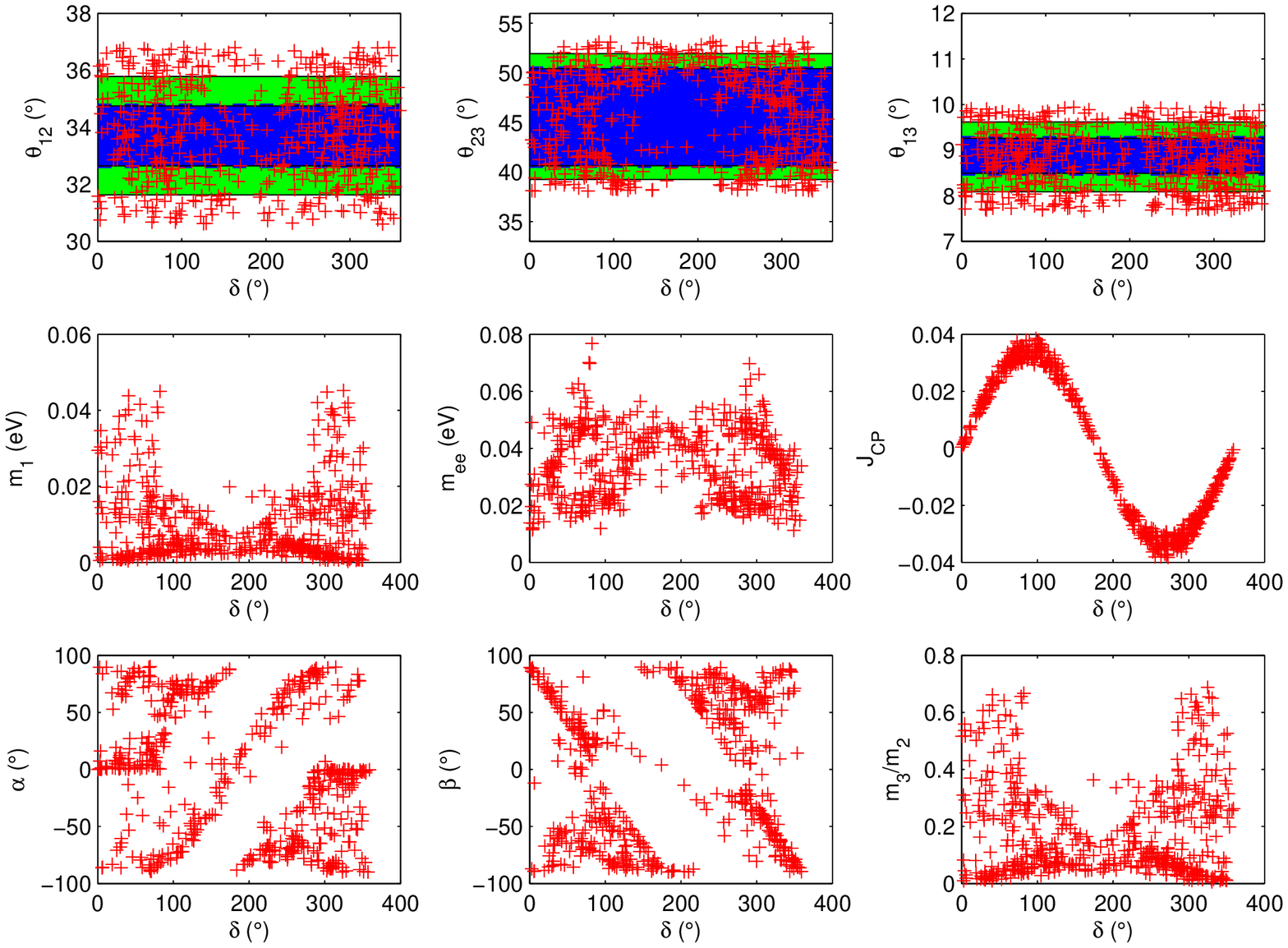}%
\caption {The correlation plots for class III(IH). The blue
horizontal bands represent the 1$\sigma$ uncertainty in
determination of $\theta_{12}, \theta_{23}$ and $\theta_{13}$ while
they plus the green horizontal bands correspond to the 2$\sigma$
uncertainty.} \label{IIIIH}
\end{figure}

\begin{figure}
 \includegraphics[scale = 0.6]{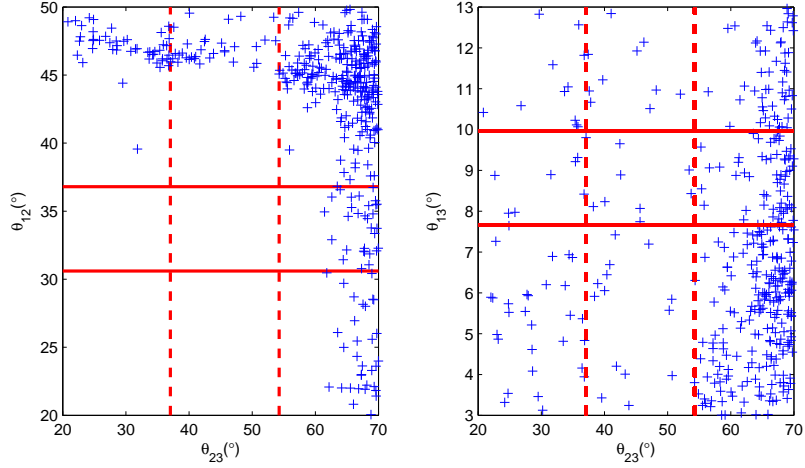}%
\caption {The correlation plots $(\theta_{23}, \theta_{12})$ and
$(\theta_{23}, \theta_{13})$ for class III(NH). The horizontal and
vertical lines respectively denote the 3$\sigma$ upper and lower
bound of $\theta_{12}$ and $\theta_{23}$} \label{IIINH}
\end{figure}

\subsection{Class III}
In the case of class III, the factorisable charged lepton matrix is
written by
\begin{equation}\begin{split}
(M_{l}^{r})^{III}=\left(\begin{array}{ccc}
0&a&b\\
a&0&c\\
b&c&d
\end{array}\right)
\end{split}\label{III}\end{equation}
where $a,b,c$ and $d$ are real number and $b,c$ are set to be
positive. The matrix $(M_{l}^{r})^{III}$ is diagonalized by the
orthogonal matrix $O_{l}$
\begin{equation}
O_{l}^{T}(M_{l}^{r})^{III}O_{l}=diag(m_{e},- m_{\mu},m_{\tau})
\label{ot}\end{equation} Here we choose $a$ as the free parameter
because $d$ has been fixed by Tr$(M_{l}^{r})$. i.e
\begin{equation}
d=m_{e}-m_{\mu}+m_{\tau} \label{IIIab}\end{equation} With the help
of other two invariant quantity Det$(M_{l}^{r})$ and
Tr$(M_{l}^{r})^{2}$, $b, c$ are determined by three charged leptonic
mass eigenvalues$(m_{e}, m_{\mu}, m_{\tau})$ and $a$
\begin{equation}
(b\pm c)^{2}=-(-m_{e}m_{\mu}+m_{e}m_{\tau}-m_{\mu}m_{\tau})-a^{2}\pm
\frac{a^{2}(m_{e}-m_{\mu}+m_{\tau})-m_{e}m_{\mu}m_{\tau}}{a}
\label{IIIac}\end{equation} Then diagonalization matrix can be
constructed as
\begin{equation}\begin{split}
(M_{l}^{r})^{III}=\left(\begin{array}{ccc}
\frac{O(11)}{N_{1}}&\frac{O(12)}{N_{2}}&\frac{O(13)}{N_{3}}\\
\frac{O(21)}{N_{1}}&\frac{O(22)}{N_{2}}&\frac{O(23)}{N_{3}}\\
\frac{O(31)}{N_{1}}&\frac{O(32)}{N_{2}}&\frac{O(33)}{N_{3}}
\end{array}\right)
\end{split}\label{III}\end{equation}
The matrix elements are given by
\begin{equation}\begin{split}
O(11)=&a
m_{e}^{-1}(bm_{e}^{-1}+ca^{-1})+bm_{e}^{-1}(m_{e}a^{-1}-m_{e}^{-1}a)\\
O(12)=&-a
m_{\mu}^{-1}(-bm_{\mu}^{-1}+ca^{-1})-bm_{\mu}^{-1}(-m_{\mu}a^{-1}+m_{\mu}^{-1}a)\\
O(13)=&a
m_{\tau}^{-1}(bm_{\tau}^{-1}+ca^{-1})+bm_{\tau}^{-1}(m_{\tau}a^{-1}-m_{\tau}^{-1}a)\\
&O(21)=bm_{e}^{-1}+ca^{-1}\\
&O(22)=-bm_{\mu}^{-1}+ca^{-1}\\
&O(23)=bm_{\tau}^{-1}+ca^{-1}\\
&O(31)=m_{e}a^{-1}-m_{e}^{-1}a\\
&O(32)=-m_{\mu}a^{-1}+m_{\mu}^{-1}a\\
&O(33)=m_{\tau}a^{-1}-m_{\tau}^{-1}a\\
\end{split}\end{equation}
with the normalized coefficients
\begin{equation}\begin{split}
N_{1}^{2}=O(11)^{2}+O(21)^{2}+O(31)^{2}\\
N_{2}^{2}=O(12)^{2}+O(22)^{2}+O(32)^{2}\\
N_{3}^{2}=O(13)^{2}+O(23)^{2}+O(33)^{2}
\end{split}\end{equation}

Repeating the previous analysis, the class III with inverted
hierarchy are now found to be acceptable by current experimental
data while the NH case are excluded. In Fig.\ref{IIIIH}, we show the
the main predictions for IH case. One can observe that no bounds are
founded on three mixing angles and Dirac CP-violating phase
$\delta$, leading to the Jarlskog rephasing invariant
$0<|J_{CP}|<0.04$. One the other hand, there is a correlation
between $\delta$ and the lightest neutrino mass $m_{3}$. One obtains
$0$eV$<m_{3}<$$0.05$eV for
$0^{\circ}<\delta<100^{\circ}(260^{\circ}<\delta<360^{\circ})$ while
$0$eV$<m_{3}<$$0.02$eV for $100^{\circ}<\delta<260^{\circ}$,
implying that both strong and mild mass hierarchy are allowed.
Interestingly, although the correlations of $(\delta, \alpha)$ and
$(\delta, \beta)$ are complicated, there exists a lower bound of
$0.01$eV on the effective Majorana neutrino mass $m_{ee}$ which is
achievable in future $0\nu\beta\beta$ experiments.

In Fig.\ref{IIINH}, we present the common space of
$(\theta_{23},\theta_{12})$ and $(\theta_{23},\theta_{13})$ for NH
case. One easily observes that parameter space of
$(\theta_{23},\theta_{12})$ is outside the $3\sigma$ allowed region
and a large corrections of $\theta_{23}$ or $\theta_{12}$ is needed.

\section{The $Z_{4}\times Z_{2}$ flavor symmetry realization}

In general, all phenomenologically viable lepton mass matrices with
with parallel texture/cofactor zeros can be realized in seesaw
models with Abelian flavor symmetry. The lepton mass matrices of
class I are equivalent to the ones with no-parallel texture zeros.
The symmetry realization of such texture structures has been
performed in Ref.\cite{zn}. Thus we only consider class II and III.
In this section, we take the first matrix pattern of class II as a
illustration. It is shown that the lepton mass matrix can be
realized based on the type-I seesaw models with the $Z_{4}\times
Z_{2}$ flavor symmetry. We take the same strategy of
Ref.\cite{minor,minor1,minor2}. In flavor basis, $M_{\nu}$ belonging
to class II is realized under $Z_{8}$ symmetry\cite{minor1}.
Different from Ref.\cite{minor1}, we build the model under the basis
where $M_{l}$ is nodiagonal. Under the $Z_{4}\times Z_{2}$ symmetry,
the three charged lepton doublets $D_{i L}=(\nu_{i L}, l_{iL})$,
three right-handed charged lepton singlets $l_{iR}$ and three
right-handed neutrinos $\nu_{iR}$ (where $i=e,\mu,\tau$) transform
as
\begin{equation}\begin{split}
\nu_{eR}\sim(\omega,1),\quad\quad\nu_{\mu
R}\sim (1,1), \quad\quad\nu_{\tau R}\sim (\omega^{2},1)\\
D_{eL}\sim (\omega,-1),\quad\quad D_{\mu L}\sim (1,-1),\quad\quad
D_{\tau L}\sim (\omega^{2},-1)\\
l_{eR}\sim (\omega^{3},-1), \quad\quad l_{\mu R}\sim
(1,-1),\quad\quad l_{\tau R}\sim (\omega^{2},-1)
\end{split}\end{equation}
where $\omega=e^{i\pi/2}$. Then, under $Z_{4}$ symmetry, the
bilinears of $\overline{D}_{iL} l_{jR}$, $\overline{D}_{iL}
\nu_{jR}$, and $\nu_{iR}^{T}\nu_{jR}$, transform respectively as
\begin{equation}
\left(\begin{array}{ccc}
-1&-i&i\\
i&1&-1\\
-i&-1&1
\end{array}\right)\quad\quad
\quad \left(\begin{array}{ccc}
1&-i&i\\
i&1&-1\\
-i&-1&1
\end{array}\right)\quad\quad\quad
\left(\begin{array}{ccc}
-1&i&-i\\
i&1&-1\\
-i&-1&1
\end{array}\right)
\label{tra}\end{equation}
To generate the fermion mass,  we need
introduce the three Higgs doublets $\Phi_{12}, \Phi_{23}, \Phi$ for
charged lepton matrix $M_{l}$, one the Higgs doublet $\Phi^{\prime}$
for Dirac neutrino mass matrix $M_{D}$ and a scalar singlet $\chi$
for the Majorana neutrino mass matrix $M_{R}$, which transform under
$Z_{4}\times Z_{2}$ symmetry as
\begin{equation}\begin{split}
\Phi_{12}\sim (\omega,1),&\quad\quad \Phi_{13}\sim (\omega^{3},1), \quad\quad\Phi\sim(1,1)\\
&\Phi^{\prime}\sim (1,-1),\quad\quad\quad\quad \chi\sim (\omega,1)
\end{split}\end{equation}
To maintain the invariant Yukawa Lagrange under the flavor symmetry
, the $\Phi_{12}$ and $\Phi_{13}$ couple to the bilnears
$\overline{D}_{eL} l_{\mu R}$ and $\overline{D}_{eL} l_{\tau R}$ to
produce the (1,2) and (1,3) nozero matrix elements in $M_{l}$ while
$\Phi$ couples to $\overline{D}_{\mu L} l_{\mu R}$
$\overline{D}_{\tau L} l_{\tau R}$ to produce the (2,2) and (3,3) no
zero matrix elements. The zero matrix elements in $M_{l}$ is
obtained because there are no appropriate scalars to generate them.
For the Dirac neutrino mass sector, there exists only one scalar
doublet $\Phi^{\prime}$ transforming invariantly under $Z_{4}$.
Therefore the $\Phi^{\prime}$ will contribute only to the (1,1),
(2,2), (3,3) no zero elements leading to a diagonal $M_{D}$. Here
the $Z_{2}$ symmetry is used to distinguish the set of scalar
doublets $(\Phi_{12}, \Phi_{13}, \phi)$ from $\Phi^{\prime}$ so that
they are respectively in charge of the mass generation of $M_{l}$
and $M_{D}$ without any crossing. In order to produce the Majorana
neutrino mass term, we introduce a complex scalar singlet $\chi$.
The $\chi$ couples to $\nu_{eR}^{T}\nu_{\tau R}$ while $\chi^{\ast}$
couples to $\nu_{eR}^{T}\nu_{\mu R}$, leading to the (1,2) and (1,3)
no zero elements in $M_{R}$. From \eqref{tra}, the $\nu_{\mu
R}^{T}\nu_{\mu R}$ and $\nu_{\tau R}^{T}\nu_{\tau R}$ is invariant
under $Z_{4}$, thus we can directly write them in the Lagrange
without needing the singlets. The zero elements in $M_{R}$ are
obtained by not introducing other scalar singlets. Therefore the
mass matrices $M_{l}$, $M_{D}$ and $M_{R}$ is given by
\begin{equation}
M_{l}\sim\left(\begin{array}{ccc}
0&\times&\times\\
\times&\times&0\\
\times&0&\times
\end{array}\right)\quad\quad\quad
M_{D}\sim\left(\begin{array}{ccc}
\times&0&0\\
0&\times&0\\
0&0&\times
\end{array}\right)\quad\quad
\quad M_{R}\sim\left(\begin{array}{ccc}
0&\times&\times\\
\times&\times&0\\
\times&0&\times
\end{array}\right)
\label{mass}\end{equation} Using the neutrino mass formula of type-I
seesaw mechanism $M_{\nu}=-M_{D}M_{R}M_{D}^{T}$, we obtain
\begin{equation}
M_{\nu}\sim\left(\begin{array}{ccc}
\Delta&\times&\times\\
\times&\times&\Delta\\
\times&\Delta&\times
\end{array}\right)
\end{equation}
Together with the $M_{l}$ in \eqref{mass}, we have realized the
leptonic mass matrices of class II with parallel texture/cofactor
zeros under $Z_{4}\times Z_{2}$ flavor symmetry. The symmetry
realization of class III can be similarly performed.

\section{Conclusion and discussion}
We have investigated the parallel texture structures with two
texture zeros in lepton mass matrix $M_{l}$ and two cofactor zeros
in neutrino mass matrix $M_{\nu}$. The 15 possible textures are
grouped into class I, II, III, and IV, where the matrices in each
class are related by means of permutation transformation and share
the same physical implications. We found only class I, II, III are
notrivial. Using the recent results of the neutrino oscillation and
cosmology experiments, a phenomenological analysis are
systematically proposed for each class and mass hierarchy. We
demonstrate the correlation plots between Dirac CP-violating phase
$\delta$, three mixing angles $\theta_{12},\theta_{23}$ and
$\theta_{13}$, the effective Majorana neutrino mass $m_{ee}$, the
lightest neutrino mass, Majorana CP-violating phase $\alpha, \beta$
and the neutrino mass ratio, leading to the predictions to be
confirmed by future experiments. A realization of the model base on
$Z_{4}\times Z_{2}$ flavor symmetry is illustrated.

Finally we would like to mention that in the spirit of Ref.
\cite{GCB, hyp}, the parallel texture structures are treated as a
natural precursor of more general cases. A systematic analysis of
all possible combinations deserves further study and will be
published in \cite{pre}.

\begin{acknowledgments}
This work is supported by the Fundamental Research Funds for the
Central Universities The author would like to thank Shu-Yuan Guo for
the helpful discussion.
\end{acknowledgments}

\end{document}